\documentclass[aps,preprint,showpacs]{revtex4}

\usepackage{graphicx}

\begin{document}

\newcommand{\be} {\begin{equation}}
\newcommand{\ee} {\end{equation}}
\newcommand{\ba} {\begin{eqnarray}}
\newcommand{\ea} {\end{eqnarray}}
\newcommand{\tr} {{\rm tr}}

\title{Statistical bounds on the dynamical production of entanglement}

\author{R\^omulo F. Abreu}
\email{romulof@cbpf.br}

\author{Ra\'ul O. Vallejos}
\email{vallejos@cbpf.br}
\homepage{http://www.cbpf.br/~vallejos}
\affiliation{ Centro Brasileiro de Pesquisas F\'{\i}sicas (CBPF), \\
              Rua Dr.~Xavier Sigaud 150, 
              22290-180 Rio de Janeiro, Brazil}

\date{\today}

\begin{abstract}
We present a random-matrix analysis of the entangling power
of a unitary operator as a function of the number of times it is 
iterated. 
We consider unitaries belonging to the circular ensembles
of random matrices (CUE or COE) applied to random 
(real or complex) non-entangled states. 
We verify numerically that the average entangling power is a 
monotonic decreasing function of time.
The same behavior is observed for the ``operator entanglement"
--an alternative measure of the entangling strength of a 
unitary.
On the analytical side we calculate the CUE operator entanglement 
and asymptotic values for the entangling power.
We also provide
a theoretical explanation of the time dependence in the 
CUE cases.
\end{abstract}



\maketitle

\section{Introduction}

In the last years many studies were devoted to the determination of entanglement 
growth laws for bipartite pure states evolving from product 
states under globally unitary dynamics, either with continuous 
\cite{furuya98,gorin03,jacquod04,znidaric05,angelo05,kubotani06,
petitjean06,pineda07}
or discrete time
\cite{miller98,miller99,laksh01,bandy02,tanaka0203,
scott03,demkowicz04,bandy04,ghose04,rossini06}.
For not too small systems, and weak couplings between subsystems, the
general qualitative picture is that of entanglement (subsystem entropy) 
growing smoothly from zero, possibly in a nonmonotonic way, until 
arriving at an asymptotic regime characterized by small oscillations 
around an equilibrium value. 
However, when we come to the quantitative level a rich phenomenology 
is discovered
\cite{gorin03,jacquod04,znidaric05,angelo05,kubotani06,petitjean06,
pineda07,laksh01,bandy02,tanaka0203,scott03,demkowicz04,bandy04,ghose04,
rossini06}.
Besides chaos or regularity at the classical level, the choice of 
parameters like subsystem dimensions, coupling strength, initial state, 
time window, etc., play also important roles in determining the law of
growth of entropy \cite{pogorzelska06}. 

In this paper we concentrate on the regime of very long times, i.e.,
after the system has relaxed to a equilibrium state.
More precisely, we are interested in the average value of the asymptotic 
entropy over a suitable distribution of initially nonentangled states.
This defines the asymptotic {\em entangling power} \cite{zanardi00} of 
the unitary dynamics.

If the classical dynamics is chaotic in the full phase space, then, 
according to the Bohigas-Giannoni-Schmit conjecture 
\cite{bohigas84,haake00},
one should expect that Random Matrix Theory will succeed
in describing the statistical features of the long time dynamics, in
particular, the distribution of asymptotic entropies. 
However, there is a much simpler statistical approach,
based on the assumption that  a typical initial state 
submitted to a ``chaotic" dynamics 
must eventually evolve into a random state, uniformly distributed on
the sphere, as far as its average properties are concerned.
This hypothesis was tested in several finite dimensional quantum maps, 
with a satisfactory quantitative agreement between theory 
and simulation \cite{bandy02,scott03,demkowicz04,vallejos06,abreu06}.

The purpose of this paper is to compare the predictions of 
Random Matrix Theory and the alluded ``Random State Theory" 
for the average asymptotic entanglement generated by a globally 
unitary map. 
In Random Matrix Theory the dynamics is explicitly introduced in the
model: an asymptotic state is generated by the repeated application of
a random unitary map to a nonentangled initial state \cite{gorin03}. 
We show that the ensemble of states generated in this way does not coincide in general with a uniform distribution on the sphere.

Our results are more conveniently stated in a language of operators: 
the entangling power \cite{zanardi00} of $U^n$, 
where $U$ is a random unitary, decreases (in average) with increasing
the discrete time $n$.
The statement continues to be true if one substitutes ``entangling
power" by ``operator entanglement" \cite{zanardi01,wang02,nielsen03},
another useful measure of the entangling abilities of a unitary 
(verified numerically, section \ref{sec3}). 

The following section (\ref{sec2}) contains the definitions and the
exact setting of the problem. 
Sections \ref{sec3} and \ref{sec4} present our numerical and analytical
results, respectively.
A brief discussion of the results is left to section \ref{sec5}.

\section{Definitions and Setting}
\label{sec2}

We restrict our analysis to the case of bipartite entanglement of 
pure states in finite dimensional Hilbert spaces.
As a measure of entanglement, we use the subsystem linear entropy.

Consider a full system divided into two subsystems, $A$ and $B$.
The dimension of the full Hilbert space $\cal{H}$ is
$d = d_A d_B$, 
with $d_A$ and $d_B$ the subsystem dimensions.
Let 
$|\psi \rangle = |\psi_A \rangle \otimes | \psi_B \rangle $ 
be a pure separable state of the full system, 
corresponding to the density matrix
$\rho = |\psi \rangle \langle \psi |$.
In general, after $n$ applications of $U$, $n \ge 1$, the new 
density matrix 
$\rho^{(n)} = U^n \rho U^{n \dagger}$ 
will not correspond to a separable state any more, due to the
increasing entanglement between the subsystems. 
This will manifest itself in a growth of the linear entropy of 
the reduced density matrices 
\be
\label{linearS}
S_L^{(n)}(|\psi\rangle)
          \equiv 1 - \tr \big[{\rho}^{(n)}_A \big]^2 
             =   1 - \tr \big[{\rho}^{(n)}_B \big]^2   \; ,  
\ee 
where
${\rho}^{(n)}_A = \tr_B \rho^{(n)}$  and  
${\rho}^{(n)}_B = \tr_A \rho^{(n)}$  
\cite{nielsen01}.
For long times, and typical $U$ and $|\psi \rangle$, the system 
comes into an equilibrium regime, where the linear entropy shows
small fluctuations around a stationary average
(see, e.g., 
\cite{bandy02,scott03,bandy04,demkowicz04,vallejos06}), 
given by
\be
\label{asympS}
S^{\infty}_L (|\psi\rangle) \equiv 
           \lim_{N \to \infty} 
           \frac{1}{N} \sum_{n=1}^N S_L^{(n)} 
               (|\psi\rangle) \; .
\ee
By doing an additional average on initial product states 
one arrives at the {\em asymptotic entangling power} of $U$: 
\be
\label{asympE}
ep^\infty(U) \equiv 
\big\langle S^\infty_L (|\psi \rangle)
\big\rangle_{|\psi\rangle=|\psi_A\rangle\otimes|\psi_B\rangle} \; .
\ee
It is also useful to consider a time-dependent entangling
power, i.e., the initial-state average of Eq.~(\ref{linearS}):
\be
\label{timedepE}
ep^{(n)}(U) \equiv 
\big\langle S^{(n)}_L (|\psi \rangle)
\big\rangle_{|\psi\rangle=|\psi_A\rangle\otimes|\psi_B\rangle} \; .
\ee
For $n=1$ this is just the entangling power of $U$.

Concerning the average over product states, we take
$|\psi_A\rangle$ and $|\psi_B\rangle$ to be independent 
random vectors, both of them either real or complex, 
uniformly drawn from the corresponding sphere
\cite{wootters90,bengtsson06}. 
In other words, the components of
$|\psi_A \rangle$ and $|\psi_B \rangle$ 
are distributed like the columns of a matrix belonging 
either to the orthogonal group (real case) or unitary group 
(complex case) (Haar measure is assumed in both cases).
There are two reasons for these choices. 
(i) They are perhaps the simplest nontrivial cases both from
a conceptual point of view \cite{wootters90} and from the 
perspective of analytical calculations. 
(ii) They will allow us to make contact with closely related 
literature 
(e.g., Refs.~\cite{zanardi00,bandy02,scott03,gorin03}). 

The problem is how to estimate $ep^\infty(U)$ for a typical 
unitary $U$. 
For ``typical unitary" we mean an operator describable  
(in a statistical sense) by any of the Circular Ensembles 
of Random Matrix Theory (RMT) \cite{mehta04} .
Accordingly we shall consider that $U$ belongs either to the
Circular Unitary Ensemble 
(CUE, unitary group with Haar measure)
or to the Circular Orthogonal Ensemble (COE), 
the latter being
the appropriate choice for unitaries displaying time 
reversal symmetry \cite{haake00}. 
This leaves us with four cases to analyze:
CUE/COE unitaries acting on random complex/real states.

In order to check that our results are not exclusive of 
the measure chosen for quantifying entangling 
strength \cite{zycz}, 
in addition to ``entangling power" we also
studied the alternative measure called 
{\em operator entanglement} \cite{zanardi01}
(also known as Schmidt strength \cite{nielsen03}),
constructed as follows.
A bipartite Hilbert space induces a bipartite structure in 
the space of its linear operators, which, equipped with the 
Hilbert-Schmidt product becomes a bipartite Hilbert space 
itself. 
Then, operators can be treated as usual vectors, and standard
measures for entanglement of states can be translated to 
operators \cite{zanardi00,wang02,bandy05,bengtsson06}. 
For instance, the linear entropy of the unitary $U$ 
reads \cite{wang02}
\be
\label{opE}
S_L(U)= 1 - \frac{1}{d_A^2 d_B^2} 
 \sum_{\; k_1,k_2,l_1,l_2=1}^{d_A}
 \sum_{\; i_1,i_2,j_1,j_2=1}^{d_B}
U_{k_1 i_1, k_2 i_2}
U_{l_1 j_1, l_2 j_2}
U_{l_1 i_1, l_2 i_2}^\ast
U_{k_1 j_1, k_2 j_2}^\ast \; ,
\ee
where the matrix elements of $U$ are related to a product 
basis, i.e., 
\be
U_{k_1 i_1, k_2 i_2} = _A \langle k_1 | 
                       _B \langle i_1 | \, U  
                    | k_2 \rangle_A 
                    | i_2 \rangle_B           \; .
\ee
Of course, by substituting $U$ by $U^n$ in the equations above 
we obtain the operator entanglement as a function of time.

\section{Numerical Results}
\label{sec3}

We start with a numerical study,
emphasizing the most interesting features, but 
postponing a deeper analysis until section \ref{sec4}.  

The main ingredients of our simulations are random states 
(real or complex) and random matrices (CUE or COE).
They were generated using the same methods as in 
Ref.~\cite{abreu06}. 
Random states are evolved by applying $n$ times
a quantum random map. 
Then, the entropy of the final state is calculated and
averaged over maps and (if necessary) over states.

\begin{figure}[htp]
\hspace{0.0cm}
\includegraphics[width=15cm]{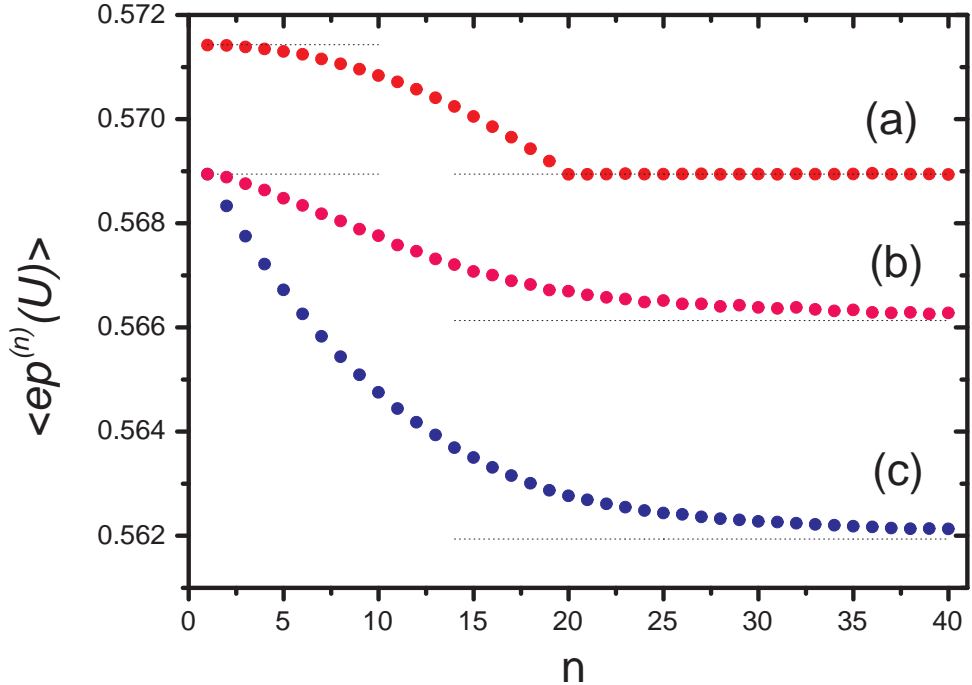}
\caption{
(a) A fixed initial separable state was chosen arbitrarily
and then propagated by $U^n$, with $U$ belonging to CUE.
Shown is the linear entropy of the evolved states averaged 
over $10^8$ CUE matrices.
(b) $10^5$ COE matrices were used to propagate an ensemble
of $10^3$ random, complex, separable initial states. 
Shown is the linear entropy averaged over COE matrices and
states. 
(c) Idem (a) but for $10^8$ COE matrices and one real state.
In all cases subsystem dimensions are $d_A=4$ and $d_B=5$.
Statistical error bars are smaller than symbol diameter.
Dotted lines correspond to analytical predictions, 
see section \ref{sec4}.}
\label{fig1}
\end{figure}

Figure~\ref{fig1} shows the entangling power of a unitary 
as a function of time, 
averaged over CUE [(a)] or COE [(b), (c)].
The cases (a,b) and (c) correspond, respectively, to complex
or real initial states. 
In cases (a) and (c), due to invariance considerations, the 
average over states is redundant and it suffices to consider 
a single state. 
For similar reasons, the cases $n=1$ represent the average 
linear entropy of standard bipartite pure states, 
either complex [(a)] or real [(b),(c)] (see section \ref{sec4}).

%

In the three cases we observe that average entanglement 
is a decreasing function of the number of iterations.
This is the opposite to what is observed in weakly coupled
maps, i.e., entropy increasing from a zero initial value.
However, we remark that our purpose is to model the equilibrium 
itself, not the initial phase of relaxation to equilibrium 
--this would require an explicit modeling of the weak coupling, 
as in Ref.~\cite{gorin03}.
The cases $n=1$ and $n \to \infty$ correspond, respectively, 
to the predictions of ``Random State Theory" and Random Matrix
Theory.
Even though these extreme cases are our main concern, 
we also analyze the regime of intermediate times because it
contains valuable information, e.g., about characteristic
times for the transition between the extremes.

\begin{figure}[htp]
\hspace{0.0cm}
\includegraphics[width=15cm]{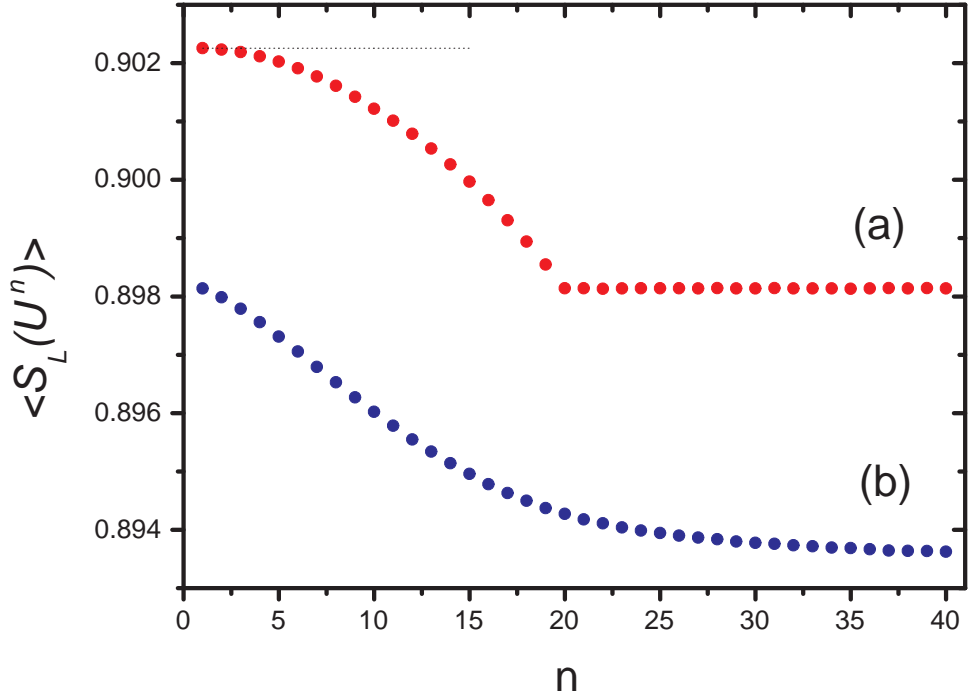}
\caption{
Shown is the linear entropy of the random operator $U^n$ 
averaged over $10^7$ realizations. (a) $U$ belongs to CUE.
(b) $U$ belongs to COE.
In both cases subsystem dimensions are $d_A=4$ and $d_B=5$.
Statistical error bars are smaller than symbol diameter.
Dotted line correspond to the analytical prediction, 
see section \ref{sec4}. }
\label{fig2}
\end{figure}

Evidently the characteristic time for saturation is the 
Heisenberg time $n_H \equiv d$ (in our simulations $d=20$). 
In the CUE case the saturation happens abruptly at $n=n_H$.
We also verified that the {\em operator entanglement} behaves
in a similar way by plotting the average linear entropy of $U^n$ 
as a function of $n$ (see Fig.~\ref{fig2}). 

%
%

Both figures exhibit a very curious characteristic: the 
asymptotic value for CUE maps coincides, within numerical
precision, with the $n=1$ value for COE. 
We shall see in the next section that, in the case of the
entangling power (Fig.~\ref{fig1}), such a coincidence 
is indeed exact, for all subsystem dimensions $d_A$ and $d_B$.

\section{Analytical Results}
\label{sec4}

The purpose of this section is to explain analytically
some of the features present in 
Fig.~(\ref{fig1}) and Fig.~(\ref{fig2}).
In some cases we shall be able to understand the global 
appearance of the entangling measures as functions of time, 
and derive quantitative expressions for some limiting values 
(indicated with dotted lines in 
Figs.~\ref{fig1} and ~\ref{fig2}).

The values for the entangling power at $n=1$ can already be 
found in the literature
\ba
\label{ep1a}
{ep^{(1)}}(U)_{(a)}     & = & 
\frac{d-(d_A+d_B)+1}{d+1} \; , \\  
\label{ep1bc}
{ep^{(1)}}(U)_{(b),(c)} & = & 
\frac{                         d^3 - 
             (d_A + d_B -4) \, d^2 -
          [3(d_A + d_B )-1] \, d   + 
           2( d_A + d_B -1)
     }
     {d(d+1)(d+3)} \; .
\ea
[Subindices (a), (b), (c) refer to each one of the cases 
depicted in Fig.~\ref{fig1}.]
The first equality above corresponds to the well known entropy 
of random complex states \cite{bengtsson06}.
The second result can be found in Ref.~\cite{gorin03}.

Concerning operator entanglement, the case $n=1$ was 
calculated by Zanardi \cite{zanardi01} for two qudits, i.e.,
$d_A=d_B$.
Using techniques to be described below, we obtained the
CUE average of Eq.~(\ref{opE}):
\be
\label{oe1a}
\left\langle {S_L (U)} \right \rangle _{\rm CUE}  = 
\frac{d^2 - (d_A^2 + d_B^2) + 1}
     {d^2 -1} \; ,
\ee
thus extending Zanardi's result to arbitrary subsystem
dimensions. 
Inserting $d_A=4$ and $d_B=5$ in this formula we obtain the
value indicated with dotted lines in Fig.~\ref{fig2}.

\subsection{Global features}

All the functions depicted in Figs.~\ref{fig1} and 
\ref{fig2} share the property of decreasing in a monotonic 
way and coming to saturation around the Heisenberg time. 
Or, equivalently, one can say that the purity 
(one minus linear entropy) grows and then saturates.
The most surprising case is CUE, because of the abrupt 
saturation at $n=n_H$. 
This behavior is similar to that of the {\em form factor} 
of the Circular Ensembles \cite{haake00}, defined as
\be
           \left \langle \left| \tr \, U^n  \right|^2 
           \right\rangle 
    \equiv \left \langle \left|     \, t_n  \right|^2 
           \right\rangle   
    =  \sum_{\alpha,\beta=1}^d 
           \left\langle 
           e^{i n (\phi_\alpha - \phi_\beta)} 
           \right\rangle                             \; ,
\ee
where the average runs over CUE or COE.
For CUE the form factor is piecewise linear:
\be
\left \langle \left| \, t_n  \right|^2  
\right\rangle_{\rm{CUE}}   
= \left\{  
\begin{array}{ll}
n & \textrm{~~ if \, $1 \le n \le d$ } \; , \\
d & \textrm{~~ if \, $      n \ge d$ } \; .
\end{array}  \right.   
\ee

The explanation for this behavior is as follows.
The form factor is a function only of the eigenvalues of $U^n$.
If $n=1$ one has the well known random matrix spectrum which 
shows strong correlations, e.g., level repulsion.
For $n>1$ the spectrum has been stretched and folded $n$ times
on the unit circle, and, when $n \gtrsim d$ 
the spectrum is almost completely uncorrelated \cite{haake00}.
Evidently, the same mechanism is responsible for the saturation 
of the entangling measures.

The similarity between purity and form factor was already noted 
by Gorin and Seligman \cite{gorin03}, who considered a continuous 
time Hamiltonian analogue of the COE case in Fig.~\ref{fig1}(c). 
Now we show that such a connection can be established rigorously
for CUE maps.
Consider either the entangling power or the operator entanglement, 
Eqs.~\ref{timedepE} and \ref{opE}, respectively;
insert the spectral decomposition for the corresponding 
unitaries. 
We recall that eigenvectors and eigenvalues are
statistically independent in the circular ensembles.
In all cases the result can be written as follows:
\be
\label{eqgen}
S(n)= 1 - \sum_{\alpha,\beta,\delta,\gamma=1}^d 
             C_{\alpha\beta\delta\gamma}
          \left\langle 
           e^{i n (\phi_\alpha + \phi_\beta
                  -\phi_\delta - \phi_\gamma)} 
          \right\rangle \; .
\ee
On the left, $S(n)$ represents any of the average entropies 
considered. 
The coefficients $C_{\alpha\beta\delta\gamma}$ contain the
average over eigenvectors and (where applicable) initial 
states.
The time dependence comes from the average over four eigenphases 
$\phi$.
Due to invariance properties of CUE and COE, the averages above 
do not depend on the particular values of the indices
$\alpha,\beta,\delta,\gamma$, but only on their
being all different, all equal, equal in pairs, etc.
Thus, one is left with the problem of evaluating a few 
nontrivial averages \cite{gorin03}:
\ba
&& \langle \exp{[\, i n (  \phi_\alpha + \phi_\beta -
                   \phi_\delta - \phi_\gamma)]} 
   \rangle \; ,\\
&& \langle \exp{[\, i n ( 2\phi_\alpha - \phi_\delta 
                               - \phi_\gamma)]} 
   \rangle \; ,\\
&& 
\label{av3}
   \langle \exp{[\, 2 i n (\phi_\alpha -\phi_\delta )]} 
   \rangle \; ,\\
&& 
\label{av4}
   \langle \exp{[\,   i n (\phi_\alpha -\phi_\delta )]} 
   \rangle \; .
\ea
For CUE, we can show that all these four averages can be 
expressed in terms of the basic form factors \cite{abreu07}
\be
\left \langle \left| \, t_n    \right|^2 \right\rangle^2 \; ,~
\left \langle \left| \, t_{2n} \right|^2 \right\rangle   \; ,~
\left \langle \left| \, t_n    \right|^2 \right\rangle   \; 
\ee
[this is immediate for averages (\ref{av3}) and (\ref{av4})].
The information we have gathered is enough for concluding 
that in CUE cases one must have
\be
\label{eq18}
S(n)= c_1 
+ c_2 \left \langle \left| \, t_n    \right|^2 \right\rangle^2
+ c_3 \left \langle \left| \, t_{2n} \right|^2 \right\rangle
+ c_4 \left \langle \left| \, t_n    \right|^2 \right\rangle \; ,
\ee
where $c_k$ are certain time-independent coefficients.

This result is not unexpected, as the same three basic functions 
above also appear in the CUE average of 
$\left|\, t_n  \right|^4$, calculated by Haake some years
ago \cite{haake00},
\be
 \left \langle \left| \, t_n    \right|^4 \right\rangle   =
2\left \langle \left| \, t_n    \right|^2 \right\rangle^2 +
 \left \langle \left| \, t_{2n} \right|^2 \right\rangle   -
2\left \langle \left| \, t_n    \right|^2 \right\rangle   \; ;
\ee
and $\langle \left|\, t_n  \right|^4 \rangle$ is structurally
very similar to the entangling measures we are considering:
\be
\left \langle \left|\, t_n  \right|^4 \right \rangle =
     \sum_{\alpha,\beta,\delta,\gamma=1}^d 
          \left\langle 
           e^{i n (\phi_\alpha + \phi_\beta
                  -\phi_\delta - \phi_\gamma)} 
          \right\rangle \; .
\ee

Whatever the exact values of $c_k$ in Eq.~(\ref{eq18}), 
the preceding analysis proves that for CUE both entangling 
power and operator entanglement 
decay quadratically and then saturate abruptly. 
(Strictly speaking the decay is piecewise quadratic;
however this effect is not perceptible in our figures,
neither is it in a plot of  
$\left \langle \left| \, t_n  \right|^4  \right\rangle$
versus $n$ \cite{abreu07}.)

The possible relationship between the form factor and the 
entangling measures in the COE cases remains a 
conjecture (Gorin-Seligman's);
the required calculations are rather more difficult and 
will not be attempted here.

\subsection{Asymptotic values}

As in the preceding subsection, the starting point for the 
calculations of asymptotic values is the general expression 
(\ref{eqgen}).
In the case of the entangling power the coefficients 
$C_{\alpha\beta\delta\gamma}$ are the result of a
double average over eigenvectors $| e_\mu \rangle$ 
and initial states $| \psi \rangle$ \cite{demkowicz04},
\be
C_{\alpha\beta\delta\gamma}= 
\Big\langle
\Big\langle 
\;
\langle e_\alpha | \psi \rangle \langle \psi | e_\delta \rangle
\langle e_\beta  | \psi \rangle \langle \psi | e_\gamma \rangle 
\, \tr_A \left[
\, \tr_B \left( | e_\alpha \rangle \langle e_\delta | \right)
   \tr_B \left( | e_\beta  \rangle \langle e_\gamma | \right) 
      \right] 
\;
\Big\rangle 
\Big\rangle \; .
\ee
The calculation of the asymptotic entangling power requires the
time average (\ref{asympS}) which washes out the eigenvalue 
dependence but enforces the pairing of indices:
   $\alpha=\delta$ and $\beta=\gamma$,
or $\alpha=\gamma$ and $\beta=\delta$.
Thus, one arrives at \cite{demkowicz04,abreu06}
\be
ep^\infty(U) =  
1 - 
\Big\langle
\Big\langle 
\;
\sum_\alpha    
\left| \langle e_\alpha | \psi \rangle \right|^4
\tr_A \left( \rho_A^{\alpha} \right)^2  -  
\sum_{\alpha \neq j} 
\left| \langle e_\alpha | \psi \rangle \right|^2
\left| \langle e_\beta  | \psi \rangle \right|^2
\left[ \tr_A \left( \rho_A^{\alpha} \rho_A^{\beta} \right) +    
       \tr_B \left( \rho_B^{\alpha} \rho_B^{\beta} \right)   
\right] 
\;
\Big\rangle 
\Big\rangle \; ,
\ee
where $\rho_A^{\alpha}$ and $\rho_B^{\alpha}$ stand for
the reduced density matrices of the eigenvector 
$|e_\alpha \rangle$:
\ba
\rho_A^{\alpha} & = &  
\tr_B |e_\alpha \rangle \langle e_\alpha |  \; , \\
\rho_B^{\alpha} & = &  
\tr_A |e_\alpha \rangle \langle e_\alpha |  \; .
\ea

In cases (a) and (c) of Fig.~\ref{fig1} the average 
over initial states is redundant.
It suffices to consider just one fixed initial product
state.
This is due to the invariance of the Haar measure with respect 
to left/right group actions, either for the
unitary [(a)] or the orthogonal [(c)] group, combined with 
the fact that local operations do not change the entropy
\cite{gorin03} (recall that the eigenvectors of CUE and 
COE are Haar distributed in the unitary and orthogonal 
groups, respectively).
So, in cases (a) and (c) we fix the initial state, 
e.g., $|\psi\rangle=|1\rangle_A \otimes |1\rangle_B$. 
In case (b) we must average 
$|\psi \rangle_A$ and $|\psi \rangle_B$ 
over their respective spheres.
In the unitary case (a) one has:
\ba
{ep^\infty}(U)_{\rm(a)}   =  1 -
& \Big\langle &
\;
\sum_\alpha 
\sum_{r,r\rq,s,s\rq} 
| U_{11       ,\alpha} |^4   \, 
  U_{rs       ,\alpha}       \,
  U_{r\rq s   ,\alpha}^\ast
  U_{r\rq s\rq,\alpha}
  U_{r    s\rq,\alpha}^\ast    -  \nonumber \\
& &  
\sum_{\alpha \neq \beta} 
\sum_{r,r\rq,s,s\rq} 
| U_{11       ,\alpha}|^2   
| U_{11       ,\beta }|^2
  U_{rs       ,\alpha}       
  U_{r\rq s   ,\alpha}^\ast
  U_{r\rq s\rq,\beta }
  U_{r    s\rq,\beta }^\ast    -   \nonumber \\
& & 
\sum_{\alpha\neq \beta} 
\sum_{r,r\rq,s,s\rq} 
| U_{11       ,\alpha}|^2   
| U_{11       ,\beta }|^2  
  U_{r    s   ,\alpha}
  U_{r    s\rq,\alpha}^\ast 
  U_{r\rq s\rq,\beta }
  U_{r\rq s   ,\beta }^\ast
\;
\Big\rangle \; .
\ea
The expression for the orthogonal case (c) is identical to
the preceding one but for $U$ a real unitary matrix.
The remaining case, (b), will be exhibited in 
Ref.~\cite{abreu07}.

In all cases, the last step is a group average of products
of eight matrix elements (not always different) belonging 
to one or two columns, i.e., 
``one- and two-vector averages of monomials of order 
eight".
For such averages we used the powerful diagrammatic 
method devised by Aubert and Lam for the unitary group
\cite{aubert03} and adapted by Braun to the orthogonal
case \cite{braun06}.
The method is based solely on the unitarity/orthogonality 
constraint and the invariance of the Haar measure under 
the group actions.
It provides explicit expressions for some integrals
and recurrence relations for others. 
As the calculations are lengthy but otherwise not 
illuminating, we skip intermediate steps \cite{abreu07} 
and jump to the final results:
\ba
\label{epinfa}
{ep^\infty}(U)_{\rm(a)} & = & 
{ep^{(1)}}(U)_{(b),(c)} \; , \\ 
\label{epinfc}
{ep^\infty}(U)_{\rm(c)} & = & 
\frac{                         d^4 -
               (d_A+d_B-13) \, d^3 -
           [12(d_A+d_B)-47] \, d^2 -
              35(d_A+d_B-1) \, d
     }
     {(d+1)(d+2)(d+4)(d+6)} \; .
\ea
The first line says that the asymptotic average entropy in
the unitary ensemble coincides with the $n=1$ value for 
COE [see Eq.~(\ref{ep1bc})], 
for all dimensions $d_A$ and $d_B$. 
This confirms the suspicion caused by examining the data in 
Fig.~\ref{fig1}.
However, we have not been able to go beyond the mere
analytical verification of the conjecture. 
The deep reasons for such a coincidence --if any-- remain
a mystery.

The second expression agrees with Gorin-Seligman's 
calculation \cite{gorin03},
who used a different method for averaging monomials over 
the orthogonal group \cite{gorin02}.

For the case (b) we obtained
\be
\label{epinfb}
{ep^\infty}(U)_{\rm(b)} = \frac{X}{Y} \; ,
\ee
with
\ba
X & = & 
                                                     d^5 +
                                              12  \, d^4 -
                                 (d_A^2+d_B^2-41) \, d^3 -
\\ \nonumber & &
                [\, 12(d_A^2+d_B^2) + 2(d_A+d_B)-30 \,] \, d^2 - 
\\ \nonumber & &
                             [\, 38(d_A^2+d_B^2)+18 \,] \, d   -
                  16(d_A^2+d_B^2)+56(d_A+d_B)-40
 \; , \\ \nonumber
Y & = &   
               (d_A+1)(d_B+1)(d+1)(d+2)(d+4)(d+6)       \; .
\ea

One of the advantages of having explicit analytical expressions 
is that we can now quantify the differences between
$n=1$ and $n \to \infty$. 
For CUE this represents the difference between the 
predictions of theories based either on random states 
or on random dynamics. 
For instance, let us consider the scaling with system
size, taking for definiteness $d_A=2$ (which
can be thought of as a definition of entangling 
power according to the multiqubit Meyer-Wallach 
measure \cite{meyerwallach}).
Then, for large $d_B$ one has 
\ba
{ep^{(1)} }(U)_{\rm(a)} - 
{ep^\infty}(U)_{\rm(a)}   & = &  \frac{1}{4 d_B^2} + \ldots \; , \\
{ep^{(1)} }(U)_{\rm(b)} - 
{ep^\infty}(U)_{\rm(b)}   & = &  \frac{1}{3 d_B^2} + \ldots \; , \\
{ep^{(1)} }(U)_{\rm(c)} - 
{ep^\infty}(U)_{\rm(c)}   & = &  \frac{7}{8 d_B^2} + \ldots \; .
\ea
So, the differences are always of second order in the system size.

\section{Conclusions}
\label{sec5}

An initially nonentangled state evolving under a globally chaotic
dynamics displays asymptotically features of randomness. 
This can be modeled by assuming that the state becomes a
completely random state, i.e., uniformly distributed on the sphere. 
Alternatively, one can assume that randomness lies in the dynamics,
and find out which is the ensemble of final states obtained in
this way.
We showed that both ensembles are different, i.e., the dynamics,
even if chaotic, does not generate ``canonical" random states.
When one includes in the model the information that states are 
generated dynamically, the ensemble-average entropy decreases
due to additional correlations among the state components. 
This shows up as the difference $n=1$ {\em vs.} $n \to \infty$.
The effect is relatively small, i.e., second-order in system size,
but it can be clearly detected in 
our figures, and might be important for small systems.

A curious byproduct of our studies is the conclusion that the 
asymptotic entangling measures for CUE operators coincide 
with the respective $n=1$ COE cases. 
Thus, the effect of explicitly including the dynamics in the
statistical modeling is equivalent to imposing a time reversal 
symmetry. 

Our results contain also a warning against excessively strong
interpretations of the Bohigas-Giannoni-Schmit conjecture,
which associates classical chaos with quantum randomness. 
Naively, one may be led to believe that ``more chaos always 
leads to more entanglement". 
However, if $U$ is classically chaotic, then $U^n$ is more chaotic, 
at least in the sense of a higher rate of phase-space mixing.
But we have seen here that higher powers of $U$ may be less 
entangling \cite{prosen02}.

\begin{acknowledgments}

We thank 
    O. Bohigas, 
    A. M. Ozorio de Almeida, 
    M. Sieber, 
    F. Toscano,
    C. Viviescas, and
    K. Zyczkowski, 
for interesting comments.
Partial financial support from 
CNPq, CAPES, PROSUL, and 
The Millennium Institute for Quantum Information 
(Brazilian agencies) is gratefully acknowledged. 
\end{acknowledgments}


\end{document}